\documentstyle[sprocl]{article}

\def\be{\begin{equation}}
\def\ee{\end{equation}}
\def\bea{\begin{eqnarray}}
\def\eea{\end{eqnarray}}

\newcommand{\s}{\smallskip}
\newcommand{\n}{\noindent}
\newcommand{\bc}{\begin{center}}
\newcommand{\ec}{\end{center}}
\newcommand{\bu}{\begin{underline}}
\newcommand{\eu}{\end{underline}}

\newcommand{\ty}{\textstyle}

\begin{document}

\title{What can we learn about Gribov copies from a formulation of QCD
in terms of gauge-invariant fields?}

\author{Kurt Haller}

\address{Department of Physics, University of Connecticut, Storrs, CT 06269}

\maketitle
\abstracts{We review the procedure by which we implemented the non-Abelian 
Gauss's law and constructed gauge-invariant fields for QCD in the temporal (Weyl) gauge. 
We point out that the operator-valued transformation that transforms gauge-dependent 
temporal-gauge fields into gauge-invariant ones has the formal structure of a gauge
transformation. We express the ``standard'' Hamiltonian for temporal-gauge QCD 
entirely in terms of gauge-invariant fields, calculate the commutation rules for these fields,
and compare them to earlier work on Coulomb-gauge QCD. We also discuss 
multiplicities of gauge-invariant temporal-gauge fields that belong to 
different topological sectors and that, in previous work, were shown to be based on the same 
underlying gauge-dependent temporal-gauge fields. We relate these multiplicities of 
gauge-invariant fields to Gribov copies.  We argue that Gribov copies 
appear in the temporal gauge, but  not when the theory is 
represented in terms of gauge-dependent fields and 
Gauss's law is left unimplemented. There {\em are} Gribov copies of the gauge-invariant 
gauge field, which can be constructed when Gauss's law is implemented.}
\section{Introduction}\label{sec:intro}
Quantum Chromodynamics can  be quantized in the temporal (or Weyl) gauge
(in which $A_0^a=0$), in a number of ways. 
When the gauge-fixing term ${\cal L}_{\mbox{gauge}}=-A_0^aG^a$  is used in the 
Lagrangian,\footnote{we 
use nonrelativistic notation, 
in which all space-time indices are subscripted and 
designate contravariant 
components of contravariant quantities 
such as $A_i^a$ or $j_i^a$, and covariant components of covariant quantities such as 
$\partial_i$. Repeated indices are summed from $1{\rightarrow}3$.} 
\begin{equation}
{\cal L}=-\frac{1}{4}F_{ij}^aF_{ij}^a+\frac{1}{2}F_{i0}^aF_{i0}^a+
j_i^aA_i^a-j_0^aA_0^a+{\cal L}_{\mbox{gauge}}-{\bar \psi}
\left(m-i{\gamma}\cdot\partial\right)\psi
\label{eq:lag}
\end{equation}
and the Dirac-Bergmann method of constrained 
quantization is applied,\cite{dirac,bergmann} 
the Lagrange multiplier field $G^a$ is incorporated 
into the time-derivative of ${\Pi}_0^a$, which, in this case is  
$D_i\Pi_i^a+j_0^a+G^a$. The presence of the Lagrange multiplier field 
assures a second-class system of constraints, and also terminates the 
chain of secondary constraints very quickly. The resulting  
Dirac commutators differ in only trivial ways 
from canonical Poisson commutators. Alternatively, 
it is possible to entirely avoid the need to consider primary constraints in the 
temporal gauge  by using the gauge-fixing term $-{\partial}_0A_0^aG^a$ 
instead of $-A_0^aG^a$, so that $-G^a$ becomes the 
momentum canonically conjugate to $A_0$. The gauge 
constraint then is ${\partial}_0A_0=0$, which, with the 
imposition of $A_0=0$ and Gauss's law at one particular time, 
implements both, the gauge condition and 
Gauss's law for all times.\cite{khymtemp}  The same procedure 
can be extended to all axial gauges
for which the gauge condition is $A_0^a+{\gamma}A_3^a=0$, where $\gamma$ is a variable real 
parameter~\cite{khgenax}. Finally, a very direct way of quantizing 
QCD in the temporal gauge is simply to set $A_0^a=0$ in the original 
Lagrangian~\cite{goldjack,jackiw,rossinp}.\s

In the process of quantizing QCD in the temporal gauge by any of the procedures outlined above, 
The Gribov ambiguity does not arise. In contrast,
the Gribov ambiguity  impedes the canonical quantization of QCD in the Coulomb gauge.
The inverse of the matrix of constraint commutators --- required for the 
implementation of the Dirac-Bergmann procedure --- 
depends on the inverse of the Faddeev-Popov operator  
$(D{\cdot}\partial)^{ab}=(\delta_{ab}{\partial^2}+g{\epsilon}^{aub}A_{n}^u{\partial_n})$.
In the Coulomb gauge,  the inverse of $(D{\cdot}\partial)^{ab}$ is not unique, and therefore the
inverse of the matrix of constraint commutators does not exist. In this way, the existence of 
Gribov copies of the Coulomb-gauge field impedes the quantization of QCD in that gauge; but
quantization of QCD in the temporal gauge is not similarly affected by this difficulty.  
It is on this basis that some 
authors have concluded that there is no Gribov ambiguity in the temporal gauge~\cite{wein}; a variety
of other arguments have also been put forth to support the conclusion that axial gauges 
have no Gribov copies~\cite{bass}.\s

It is significant that when QCD is quantized in the temporal gauge, Gauss's law is not implemented ---
 the imposition  of the non-Abelian Gauss's law and the construction of gauge-invariant fields 
remains to be addressed. If it were possible to use the Dirac-Bergmann procedure consistently
to quantize QCD in the 
Coulomb gauge, the implementation of Gauss's law would be an inevitable consequence --- it 
would appear as one of the Dirac-Bergmann secondary constraints (specifically from the 
Poisson bracket of the total Hamiltonian with the gauge condition $\partial_iA_i^a=0$). 
The procedure would then lead to a Hamiltonian expressed in terms of unconstrained, independent 
field variables. As it is, however, the quantization of QCD in the Coulomb gauge is very 
problematical for a number of reasons --- 
 Gribov copies and ordering ambiguities of non-commuting operators, among them.\s

In our work, we have shown how the temporal-gauge formulation of QCD 
can be brought into compliance with Gauss's law, by
explicitly constructing a set of state vectors 
that are annihilated by the Gauss's law operator  
\be
{\hat {\cal G}}^a({\bf{r}})=\overbrace{\partial_i\Pi^a_i({\bf{r}})+
\underbrace{gf^{abc}A^b_i({\bf{r}}){\Pi}^c_i({\bf{r}})}_{J^a_0({\bf{r}})}}^{D_i\Pi^a_i
({\bf{r}})\,{\equiv}\,{\cal G}^a({\bf{r}})}+j^a_0({\bf{r}}),  
\label{eq:GL}
\ee
where $j^a_0({\bf{r}})$ is the quark color-charge density 
$j^a_0({\bf{r}})=g\psi^{\dagger}({\bf{r}})\frac{{\lambda}^a}{2} \psi({\bf{r}});$ 
$J^a_0({\bf{r}})$ is the glue color-charge density as well as part of the covariant 
derivative of ${\Pi}^a_i$, which is the negative chromoelectric field and also is canonically 
conjugate to the gauge field $A^a_i$. The state vectors that are annihilated by ${\hat {\cal G}}^a$
comprise a space in which the time evolution 
of the theory takes place. In this way, the validity of Gauss's law within the framework of 
temporal-gauge QCD is assured.

\section{Gauss's Law and Gauge-Invariant Quark and Gluon Fields}\label{sec:gl}
Gauss's law in temporal-gauge QCD is implemented by constructing state vectors 
${\Psi}\,|{\phi}\rangle$ for which 
$\{\,\partial_{i}\Pi^a_{i}({\bf{r}}) + J_{0}^{a}({\bf{r}})\,\}{\Psi}\,|{\phi}\rangle =0,$ 
where $|{\phi}\rangle$ is the state annihilated by the Abelian part of $D_i\Pi^a_{i}$, so that
$\partial_i\Pi^a_{i}|{\phi}\rangle=0$ and $\Psi$ is an operator that essentially converts 
$D_i\Pi^a_{i}$ to $\partial_i\Pi^a_{i}$ within the space of $|{\phi}\rangle$ states~\cite{BCH1,CBH2}. 
$\Psi$ must obey the operator-valued differential equation 
\be
 \{\,\partial_{i}\Pi^a_{i}({\bf{r}}) + J_{0}^{a}({\bf{r}})\;\}\;
{\Psi}\,|{\phi}\rangle=
{\Psi}\,\partial_{i}\Pi^a_{i}({\bf{r}})\,|{\phi}\rangle
\label{eq:de1}
\ee
\be
\mbox{or}\;\;\;[\,\partial_{i}\Pi^a_{i}({\bf{r}}),\,{\Psi}\,]=-J_{0}^{a}({\bf{r}})\,
{\Psi}\,+\,B_Q^{a}({\bf{r}}),
\label{eq:de2}
\ee
where $B_{Q}^{a}({\bf{r}})$ is an 
operator that has
$\partial_{i}\Pi^a_{i}({\bf{r}})$ on its extreme right so that $B_{Q}^{a}({\bf{r}})|{\phi}\rangle=0$.
The solution of Eq. (\ref{eq:de2}) has been given as~\cite{CBH2}
\be\Psi={\|}\,\exp({\cal{A}})\,{\|}\;,
\label{eq:Psi}
\ee  in which 
 \be{\cal{A}}=
i{\int}d{\bf{r}}\;\overline{{\cal{A}}_{i}^{\gamma}}({\bf{r}})\;
\Pi_i^{\gamma}({\bf{r}}),
\label{eq:rf}
\ee 
and $\overline{{\cal{A}}_{i}^{\gamma}}({\bf{r}})$ is the 
\begin{underline}{Resolvent Field}\end{underline}. The resolvent field is central to 
this work, and largely determines
the properties of gauge-invariant fields and of QCD represented in terms of 
those gauge-invariant fields. The $||\;\;||$-ordered product orders $\Psi$ so that  
all functionals of $A^a_i$ are 
\begin{underline}{to the left of}\end{underline} all functionals of $\Pi^b_j\,;$ 
this ordering  is 
important because, in the temporal gauge, 
\be
[A^a_i({\bf{x}}),\Pi^b_j({\bf{y}})]=
i\delta_{ab}\delta_{ij}{\delta}({\bf{x}}-{\bf{y}}).
\label{eq:ccr}
\ee
We have shown that the 
resolvent field obeys the integral equation~\cite{CBH2}
\be
{\int}d{\bf r}\overline{{\cal A}_{j}^{\gamma}}({\bf r})V_{j}^{\gamma}({\bf r})=
\sum_{\eta=1}^\infty
{\textstyle\frac{ig^\eta}{\eta!}}{\int}d{\bf r}\;
\!\left\{\,\psi^{\gamma}_{(\eta)j}({\bf{r}})+
f^{\vec{\alpha}\beta\gamma}_{(\eta)}\,
{\cal{M}}_{(\eta)}^{\vec{\alpha}}({\bf{r}})\,
\overline{{\cal{B}}_{(\eta) j}^{\beta}}({\bf{r}})\,\right\}\;
\!\!V_{j}^{\gamma}({\bf r})
\label{eq:inteq}
\ee
in which ${\cal{M}}_{(\eta)}^{\vec{\alpha}}({\bf{r}})=\prod_{m=1}^\eta\;
\overline{{\cal Y}^{\alpha[m]}}({\bf{r}})$, $\overline{{\cal Y}^{\alpha}}({\bf r})=
{\textstyle \frac{\partial_{j}}{\partial^{2}}\overline{{\cal A}_{j}^{\alpha}}({\bf r})}\,,$
and  $\overline{{\cal B}_{(\eta) i}^{\beta}}({\bf r})=
a_i^{\beta}({\bf r})+\,
\left(\,\delta_{ij}-{\textstyle\frac{\eta}{(\eta+1)}}
{\textstyle\frac{\partial_{i}\partial_{j}}{\partial^{2}}}\,\right)
\overline{{\cal A}_{i}^{\beta}}({\bf r})\,,$ 
thus producing the nonlinearity in Eq. (\ref{eq:inteq}). 
$\psi^{\gamma}_{(\eta)i}({\bf{r}})$ 
is taken as the ``source'' term
$\psi^{\gamma}_{(\eta)i}({\bf{r}})= \,(-1)^{\eta-1}\,
f^{\vec{\alpha}\beta\gamma}_{(\eta)}\,
{\cal{R}}^{\vec{\alpha}}_{(\eta)}({\bf{r}})\;
{\cal{Q}}_{(\eta)i}^{\beta}({\bf{r}})\;,$ 
where ${\cal{R}}^{\vec{\alpha}}_{(\eta)}({\bf{r}})=
\prod_{m=1}^\eta{\cal{X}}^{\alpha[m]}({\bf{r}})$,
${\cal{X}}^\alpha({\bf{r}}) =
\,{\textstyle\frac{\partial_j}{\partial^2}}A_j^\alpha({\bf{r}})\,$, and 
${\cal{Q}}_{(\eta)i}^{\beta}({\bf{r}}) =
[\,a_i^\beta ({\bf{r}})+
{\textstyle\frac{\eta}{(\eta+1)}}\,x_i^\beta({\bf{r}})\,]\;;$
$a_i^{\gamma}({\bf r})$ and $x_i^{\gamma}({\bf r})$ are the transverse and longitudinal parts
of the gauge field respectively, and $f^{\vec{\alpha}\beta\gamma}_{(\eta)}$ denotes
the chain of structure constants
\be
f^{\vec{\alpha}\beta\gamma}_{(\eta)}=f^{\alpha[1]\beta b[1]}\,
\,f^{b[1]\alpha[2]b[2]}\,f^{b[2]\alpha[3]b[3]}
f^{b[\eta-2]\alpha[\eta-1]b[\eta-1]}f^{b[\eta-
1]\alpha[\eta]\gamma}\;.
\ee
It is an interesting fact that Eq. (\ref{eq:inteq}) for the SU(2) case takes a form strongly 
reminiscent of a finite gauge transformation for fields in an adjoint representation
of SU(2)~\cite{CBH2}.\s 

${\hat {\cal G}}^a$ and ${\cal G}^a=D_i\Pi_i^a$ are unitarily equivalent, and are related by 
\be
\hat{\cal{G}}^a({\bf{r}})={\cal{U}}_{\cal{C}}\,
{\cal{G}}^a({\bf{r}})\,{\cal{U}}^{-1}_{\cal{C}}
\label{eq:unit}
\ee
where $\,{\cal{U}}_{\cal{C}}=e^{{\cal C}_{0}}
e^{\bar {\cal C}}$ and  where $\,{\cal C}_{0}=i\,\int d{\bf{r}}\,
{\textstyle {\cal X}^{\alpha}}({\bf r})\,j_{0}^{\alpha}({\bf r})
\;\mbox{and}\;
{\bar {\cal C}}=i\,\int d{\bf{r}}\,
\overline{{\cal Y}^{\alpha}}({\bf r})\,j_{0}^{\alpha}({\bf r})\;.$
With this unitary equivalence, ${\cal G}^a$ can be used to represent ${\hat {\cal G}}^a$ in a 
new representation, which we will call the ${\cal N}$ representation. 
In this representation, the quark field $\psi$ is gauge-invariant 
because it commutes with ${\cal G}^a$.
This unitary equivalence can be used to construct gauge invariant operator-valued quark fields
in the original representation (which we call the ${\cal C}$ representation):
\be
{\psi}_{\sf GI}({\bf{r}})={\cal{U}}_{\cal C}\,
\psi({\bf{r}})\,{\cal{U}}^{-1}_{\cal C}\;\;\mbox{and}\;\;
{\psi}_{\sf GI}^\dagger({\bf{r}})={\cal{U}}_{\cal C}\,
\psi^\dagger({\bf{r}})\,{\cal{U}}^{-1}_{\cal C}
\label{eq:va}
\ee
\be\mbox{or, equivalently,}\;\;\;
{\psi}_{\sf GI}({\bf{r}})=V_{\cal{C}}({\bf{r}})\,\psi ({\bf{r}})
\;\;\;\mbox{\small and}\;\;\;
{\psi}_{\sf GI}^\dagger({\bf{r}})=
\psi^\dagger({\bf{r}})\,V_{\cal{C}}^{-1}({\bf{r}}),
\label{eq:vb}
\ee
\be
\mbox{where}\;\;V_{\cal{C}}({\bf{r}})=
\exp\left(\,-ig{\overline{{\cal{Y}}^\alpha}}({\bf{r}})
{\textstyle\frac{\lambda^\alpha}{2}}\,\right)\,
\exp\left(-ig{\cal X}^\alpha({\bf{r}})
{\textstyle\frac{\lambda^\alpha}{2}}\right)\;.
\label{eq:vc}
\ee
The transformation of $\psi$ to the
gauge-invariant ${\psi}_{\sf GI}({\bf{r}})=V_{\cal{C}}({\bf{r}})\,\psi ({\bf{r}})$
 {\em  has the formal structure of a gauge transformation\,,}
because $V_{\cal C}$ can be written as 
\be
V_{\cal C}=\exp\left[-ig{\cal Z}^\alpha
{\textstyle\frac{{\lambda}^\alpha}{2}}\right]=
\exp\left[-ig{\overline {\cal Y}^\alpha}
{\textstyle\frac{{\lambda}^\alpha}{2}}\right]\,
\exp\left[-ig{\cal X}^\alpha
{\textstyle\frac{{\lambda}^\alpha}{2}}\right]\,.
\label{eq:vz}
\ee
However, in fact, under a gauge transformation, $V_{\cal C}$ and $\psi$ \bu{both}\eu~
are transformed, so that 
\be
{\psi}{\rightarrow}\exp(-i{\omega}^{\gamma}\,
\textstyle{\frac{{\lambda}^{\gamma}}{2}})\psi\;\;\;
\mbox{and}\;\;\;V_{\cal C}{\rightarrow}
V_{\cal C}\exp(i{\omega}^{\gamma}\,\textstyle{\frac{{\lambda}^{\gamma}}{2}})
\label{eq:vztrans}
\ee
and ${\psi}_{\sf GI}({\bf{r}})$ remains strictly gauge invariant. Furthermore, the formal
similarity of $V_{\cal C}$ to a gauge transformation enables us to construct gauge-invariant 
gluon fields 
\be
\,A_{{\sf GI}\,i}^{b}({\bf{r}})\,{\textstyle\frac{\lambda^b}{2}}
=V_{\cal{C}}({\bf{r}})\,[\,A_{i}^b({\bf{r}})\,
{\textstyle\frac{\lambda^b}{2}}\,]\,
V_{\cal{C}}^{-1}({\bf{r}})\,+
{\textstyle\frac{i}{g}}\,V_{\cal{C}}({\bf{r}})\,
\partial_{i}V_{\cal{C}}^{-1}({\bf  r})\;,
\label{eq:ag1}
\ee
\be
\mbox{which is equivalent to }\;\;A_{{\sf GI}\,i}^{b}({\bf{r}})\,=
A\,_{T\,i}^b ({\bf{r}}) +
[\delta_{ij}-{\textstyle\frac{\partial_{i}\partial_j}
{\partial^2}}]\,\overline{{\cal A}_{i}^b} ({\bf{r}}),
\label{eq:ag2}
\ee
so that $A_{{\sf GI}\,i}^{b}$ turns out to be transverse ---- the sum of the 
transverse parts of the
gauge field and the resolvent field.\footnote{Iterative expansions of these 
gauge-invariant quark and gluon fields agree with the iterative representations of these fields
given by Lavelle and McMullan~\cite{LMNP,LMPR}.}\s

We have been able to express the QCD Hamiltonian entirely in terms of gauge-invariant field 
operators~\cite{BCH3,CHQC,KHG}. For that purpose, we have defined gauge-invariant versions of 
$\Pi_i^a$ and of the Gauss's law operator. These quantities are 
${\Pi}_{{\sf GI}\,i}^b={\cal R}_{bd}\Pi^{d}_{i}$ and 
${\cal{G}}_{\sf GI}^b={\cal R}_{bd}{\cal G}^d$ respectively, 
where ${\cal R}_{bd}={\textstyle\frac{1}{2}}{\sf Tr}[
V_{\cal{C}}^{-1}{\lambda^b}V_{\cal{C}}{\lambda^d}].$
This QCD  Hamiltonian (in the ${\cal N}$ representation, in which $\psi$ is the gauge-invariant form
of the quark field) is
\begin{eqnarray}
&&\!\!\!\!\!\!\!\!\!{H}_{\cal N}=\int d{\bf r}\left\{\,{\textstyle \frac{1}{2}}
\Pi^{a}_{{\sf GI}\,i}({\bf r})\,\Pi^{a}_{{\sf GI}\,i}({\bf r})
+  {\textstyle \frac{1}{4}} F_{{\sf GI}\,ij}^{a}({\bf r})
F_{{\sf GI}\,ij}^{a}({\bf r})+\right.\nonumber\\ \nonumber\\ &&\;\;\;\;\;\left.{\psi^\dagger}({\bf r})
\left[\,\beta m-i\alpha_{i}\left(\,\partial_{i}
-igA_{{\sf GI}\,i}^{a}({\bf r})
{\textstyle\frac{\lambda^\alpha}{2}}\,\right)\,\right]
\psi({\bf r})\right\}+{\tilde{H}}_{LR}+{\tilde{H}}_{\cal G} 
\label{eq:HN}
\end{eqnarray} where
\be
{\tilde{H}}_{LR}=\int d{\bf r}\left({\textstyle\frac{1}{2}}J_{0\,({\sf GI})}^{a\,\dagger}
\frac{1}{\partial^2}{\cal K}_0^a({\bf r})+
{\textstyle\frac{1}{2}}\,{\cal K}_0^a({\bf r})\frac{1}
{\partial^2}J_{0\,({\sf GI})}^{a}
-{\textstyle\frac{1}{2}}{\cal K}_0^a({\bf r})\frac{1}
{\partial^2}{\cal K}_0^a({\bf r})\right)
\label{eq:HLR}
\ee 
\be
\mbox{and}\;\;H_{\cal G}=-{\textstyle\frac{1}{2}}\int d{\bf r}\left[{\cal G}_{\sf GI}^{a}({\bf r})
\frac{1}{\partial^2}{\cal K}_0^a({\bf r})+
{\cal K}_0^a({\bf r})\frac{1}{\partial^2}{\cal G}_{\sf GI}^{a}({\bf r})\right].
\label{eq:HG}
\ee
$J_{0\,({\sf GI})}^{a}$ is the gauge-invariant glue color charge density 
$J_{0\,({\sf GI})}^{a}=gf^{abc}A_{{\sf GI}\,i}^{b}{\Pi}_{{\sf GI}\,i}^c$,  
$F_{{\sf GI}\,ij}^{a}$ is the gauge-invariant chromomagnetic field in which $A_{{\sf GI}\,i}^{a}$
replaces the gauge-dependent $A_{i}^{a}$, and ${\cal K}_0^a$ is
a gauge-invariant effective quark color charge density, given by  
\be
\left(\delta_{ab}+gf^{aub}A_{{\sf GI}\,i}^{u}
\frac{\partial_i}{\partial^2}\right){\cal K}_0^b=-j_0^a\;\;
\mbox{for}\;\;j_0^a=g\psi^{\dagger}\textstyle{\frac{\lambda^a}{2}}\psi.
\label{eq:Kj}
\ee
The relation of ${H}_{\cal N}$ to the Coulomb-gauge formulation of QCD is of particular interest. 
We first note the following commutation relations:~\cite{KHG}
\be
\left[{\Pi}_{{\sf GI}\,j}^b({\bf y})\,,A^a_{{\sf GI}\,i}({\bf{x}})\right]=
-i\{\delta_{ab}{\delta}({\bf y}-{\bf x})-\partial_j{\cal D}^{bh}({\bf y},{\bf x})
\stackrel{\leftarrow}D_i^{\,ha}\!\!\!({\bf x})\}
\label{eq:capi}
\ee
where ${\cal D}^{bh}({\bf y},{\bf x})$ designates the formal series representation of the 
inverse of the Faddeev-Popov operator, and the arrow indicates action to the left; and 
\be
\left[{\Pi}_{{\sf GI}\,i}^{\alpha}({\bf x})\,,{\Pi}_{{\sf GI}\,j}^{\beta}({\bf y})\right]=
ig\{{\partial_i}{\cal D}^{{\alpha}h}({\bf x},{\bf y}){\epsilon}^{h\gamma\beta}
{\Pi}_{{\sf GI}\,j}^{\gamma}({\bf y})-
{\partial_j}{\cal D}^{{\beta}h}({\bf y},{\bf x}){\epsilon}^{h\gamma\alpha}
{\Pi}_{{\sf GI}\,i}^{\gamma}({\bf x})\}.
\label{eq:cpipi}
\ee
These commutation rules agree with those given by Schwinger for QCD in the Coulomb gauge, 
{\em modulo} operator order in corresponding expressions~\cite{schwingera}. The Hamiltonian
$({H}_{\cal N}-{H}_{\cal G})$ also has many features in common with 
the Hamiltonian in earlier work on Coulomb-gauge QCD by 
Christ and Lee~\cite{christlee}, by Creutz {\em et. al.}\cite{creutz}, and by Gervais and 
Sakita~\cite{sakita}.\footnote{Note, however, that the chromoelectric field in these works are 
taken to be purely transverse.} ${H}_{\cal N}$ however is \bu{not}\eu~ the Coulomb-gauge 
Hamiltonian.  ${H}_{\cal G}$ ``remembers'' that the Hamiltonian is in the temporal gauge; but, 
as we have shown elsewhere~\cite{KHG}, ${H}_{\cal G}$ can neither affect the time evolution of 
state vectors within the space of states annihilated by the Gauss's law operator, nor can 
time evolution transport state vectors out of that space. The situation in QCD is, therefore, 
remarkably like that in QED. When Gauss's law is imposed and the Hamiltonian is expressed in 
terms of physical variables, the time-evolution mediated by the temporal-gauge Hamiltonian is 
essentially that expected in the Coulomb-gauge (with the proviso that the latter is well-defined
for QED but much less well-defined for QCD), but, because of the inclusion of ${H}_{\cal G}$ (or 
its QED equivalent), the Hamiltonian 
remains in its original (in this case, temporal) gauge~\cite{khqedtemp,khelqed}.\s

 Even though
${H}_{\cal N}$ is a functional of gauge-invariant --- physical --- operator-valued variables only,
complete control over its properties is retained, because all gauge-invariant 
quantities  are defined explicitly in 
terms of the gauge-dependent $A_i^a$ and $\Pi_j^b$, whose commutation rules are the 
canonical ones given in Eq. (\ref{eq:ccr}). The suggestive position in Eq. (\ref{eq:HLR})
occupied by the effective color-charge density ${\cal K}_0^a$ makes it important to 
investigate its properties in depth.

\section{Topology and the Gribov Ambiguity}\label{sec:gribov}
In earlier work, we have represented SU(2) versions of 
 $\overline{{\cal{A}}_{i}^{\gamma}}({\bf{r}})$ and 
$A_i^{\gamma}({\bf r})$ as $c$-number functions of spatial variables --- as second-rank tensors 
in the combined spatial and SU(2) indices $i$ and $\gamma$ respectively, 
so that, except in so far as 
the forms of $\overline{{\cal{A}}_{i}^{\gamma}}({\bf{r}})$ and 
$A_i^{\gamma}({\bf r})$ must reflect this second-rank 
tensor structure, they are isotropic functions of the position~\cite{HCC}. With this representation
we obtain
\be
\overline{{\cal {A}}_i^{\gamma}}\,^L({\bf r})=\frac{1}{g}\left[\left({\delta}_{{i}\,{\gamma}}\,-
\frac{r_i\,r_{\gamma}}{r^2}\right)\frac{\overline{\cal{N}}{(r)}}{r}+\frac{r_i\,r_{\gamma}}{r^2}\,
\overline{\cal{N}}^{\,\prime}{(r)}\right],
\label{eq:crfl}
\ee
\be
\overline{{\cal {A}}_i^{\gamma}}\,^T({\bf r})={\delta}_{{i}\,{\gamma}}\,{\varphi}_{A}(r)+
\frac{r_i\,r_{\gamma}}{r^2}\,{\varphi}_{B}(r)+{\epsilon}_{i{\gamma}n}
\frac{r_n}{r}\,{\varphi}_{C}(r),\;\;\mbox{as well as}
\label{eq:crft}
\ee
\be
A_i^{\gamma}\,^L({\bf r})=\frac{1}{g}\left[\left({\delta}_{{i}\,{\gamma}}\,-
\frac{r_i\,r_{\gamma}}{r^2}\right)\frac{{\cal{N}}{(r)}}{r}+\frac{r_i\,r_{\gamma}}{r^2}\,
{\cal{N}}^{\,\prime}{(r)}\right]\;\;\mbox{and}
\label{eq:cfl}
\ee
\be
A_i^{\gamma}\,^T({\bf r})={\delta}_{{i}\,{\gamma}}\,{\cal T}_{A}(r)+
\frac{r_i\,r_{\gamma}}{r^2}\,{\cal T}_{B}(r)+
{\epsilon}_{i{\gamma}n}\frac{r_n}{r}\,{\cal T}_{C}(r)\,,
\label{eq:cft}
\ee
where, to preserve transversality in the transverse fields, 
$$(r^2{\varphi}_B)'+r^2({\varphi_A})'=0\;\;\mbox{and}\;\; 
 (r^2{\cal T}_B)^{\prime}+r^2({\cal T}_A)^{\prime}=0.$$ 
With this representation, $A_{{\sf GI}\,i}^{\gamma}({\bf{r}})$ becomes
\begin{eqnarray}
&&A_{{\sf GI}\,i}^{\gamma}({\bf{r}})=\frac{1}{gr}\left\{\epsilon_{i\,\gamma\,n}\frac{r_n}{r}
\left[{\cos}(\overline{\cal{N}}+{\cal{N}})-1+{\cal{N}}\sin(\overline{\cal{N}}+{\cal{N}})\right]
+\left({\delta}_{i\,\gamma}-\frac{r_ir_{\gamma}}{r^2}\right)\times\right.\nonumber\\
&&\left.\times\left[{\cal{N}}{\cos}(\overline{\cal{N}}+
{\cal{N}})-{\sin}(\overline{\cal{N}}+{\cal{N}})\right]
-\frac{r_ir_{\gamma}}{r}\frac{d\overline{\cal{N}}}{dr}\right\}\nonumber \\
&&+{\cal T}_A\left\{
\left({\delta}_{i\,\gamma}-\frac{r_ir_{\gamma}}{r^2}\right)
{\cos}(\overline{\cal{N}}+{\cal{N}})
+\epsilon_{i\,\gamma\,n}\frac{r_n}{r}{\sin}(\overline{\cal{N}}+
{\cal{N}})\right\}+\frac{r_ir_{\gamma}}{r^2}
\left({\cal T}_A+{\cal T}_B\right)\nonumber \\
&&+{\cal T}_C\left\{\epsilon_{i\,\gamma\,n}\frac{r_n}{r}{\cos}(\overline{\cal{N}}+{\cal{N}})
-\left({\delta}_{i\,\gamma}-\frac{r_ir_{\gamma}}{r^2}\right)
{\sin}(\overline{\cal{N}}+{\cal{N}})\right\}.
\label{eq:AGIsub}
\end{eqnarray}
When Eqs. (\ref{eq:crfl} -\ref{eq:cft}) are substituted into the 
integral equation --- Eq. (\ref{eq:inteq}) --- 
for the SU(2) case, it becomes a nonlinear differential equation in which 
${\overline{\cal{N}}{(r)}}$ is the dependent function of $r$, and the other functions are source 
terms. When the independent variable is changed to $u=\ln(\ty\frac{r}{r_0})$, this equation becomes
\bea
\frac{d^2\,\overline{\cal{N}}}{du^2}&+&\frac{d\,\overline{\cal{N}}}{du}+
2\left[{\cal{N}}{\cos}(\overline{\cal{N}}+{\cal{N}})-
{\sin}(\overline{\cal{N}}+{\cal{N}})\right]\nonumber \\
&+&2gr_0\exp(u)\left\{{\cal T}_A\left[{\cos}(\overline{\cal{N}}+{\cal{N}})-1\right]-{\cal T}_C\,
{\sin}(\overline{\cal{N}}+{\cal{N}})\right\}=0.
\label{eq:nueq}
\eea
Since ${\overline{\cal{N}}{(r)}}$ must be bounded everywhere in $0{\leq}r<\infty$, 
${\overline{\cal{N}}{(u)}}$ must be bounded everywhere in $-\infty<r<\infty$. Studies 
have shown that, for the \bu{identical}\eu~ source terms ${\cal{N}}$, ${\cal T}_A$, and ${\cal T}_C$,
there are a number of different solutions of this equation, and they define different topological 
sectors of the gauge theory~\cite{HCC}. Although 
$A_{{\sf GI\,}i}^{\gamma}\,({\bf r})$ is invariant to ``small''
gauge transformations, which are continuously deformable to the identity, they can exist in
different topological sectors which are connected by ``large'' gauge transformations. 
The different topological sectors manifested when 
Eq. (\ref{eq:nueq}) is solved, reflect the existence of these large gauge transformations.
When the gauge field vanishes, $i.\,e,$ $A_i^{\gamma}({\bf{r}})=0$, 
the differential equation reduces to the autonomous equation
\be
\frac{d^2\,\overline{\cal{N}}}{du^2}+\frac{d\,\overline{\cal{N}}}{du}-
2{\sin}(\overline{\cal{N}})=0,
\label{eq:nueqzero}
\ee
 which is the equation for a damped pendulum with large amplitude
vibrations (also: Gribov equation) \bu{with the following {\em  caveat}:}\eu~
For the pendulum, $u$ is the time, and $\overline{\cal{N}}$ for that case is bounded 
for $0{\leq}u<\infty$.
The equation then has multiple solutions, corresponding to multiple turns and 
swings executed before the pendulum comes
to rest at equilibrium.
For the gauge theory, $u=\ln(\ty\frac{r}{r_0})$, and $\overline{\cal{N}}$ for 
that case is bounded for $-\infty<u<\infty$. 
The equation now does \bu{not}\eu~have multiple solutions and there are no 
Gribov copies~\cite{gribovb}. \s

Gribov discussed the case of a transverse SU(2) gauge field
 \be
{\sf A}_i={\sf A}^c_i\frac{{\tau}^c}{2}
\;\;\mbox{with}\;\;{\sf A}^c_i={\epsilon}^{ijc}\frac{r_j}{r^2}f(r),
\label{eq:grib1}
\ee
gauge-transformed to   
\be
{\sf A}^{\prime}_i=U{\sf A}_iU^{-1}+iU\partial_iU^{-1}          
\label{eq:grib_f}\;\;
\mbox{where}\;\;U=\exp\left(-i{\phi}(r)\frac{{\vec r}\cdot
{\vec \tau}}{2r}\right).
\label{eq:grib2}
\ee
The condition that ${\sf A}^{\prime}_i$ is also transverse, leads to 
\be
\frac{d^2\,\phi}{du^2}+\frac{d\,\phi}{du}-2{\sin}(\phi)\left(1-f(u)\right)=0.
\label{eq:grib3}
\ee
$f(u)$ keeps Eq. (\ref{eq:grib3}) from being autonomous, and it has 
multiple solutions for the same $f(u)$, as does our Eq. (\ref{eq:nueq}). We draw the 
following conclusions from these observations: \s

\n
$\bullet$ When QCD is quantized in the Coulomb gauge, 
the quantization procedure is impeded by the existence of Gribov copies, 
because the non-uniqueness of the inverse of the faddeev-Popov 
operator prevents the inversion of the Dirac constrained
commutator matrix.\s

\n
$\bullet$ When the quantization is carried out in the temporal 
gauge, no impediments to
the inversion of the commutator matrix arise, and the 
procedure can easily be carried out in a variety of ways. But, in contrast to the Coulomb gauge, 
Gauss's law is not implemented in the process. \s

\n
$\bullet$ Gribov copies arise in the temporal gauge when Gauss's law is 
implemented. It is the imposition
of gauge invariance and the imposition of Gauss's law that generates Gribov copies in QCD. This is 
consistent with the theorem proven by Singer~\cite{Sing}.

\section*{Acknowledgments}
This research was supported by the Department of Energy
under Grant No. DE-FG02-92ER40716.00.

\end{document}